\documentclass[nofootinbib,onecolumn,showpacs,preprintnumbers,amsmath,amssymb]{revtex4}
\def\bea{\begin{eqnarray}}
\def\eea{\end{eqnarray}}
\newcommand{\be}{\begin{equation}}
\newcommand{\ee}{\end{equation}}

\newcommand{\bn}{\begin{eqnarray}}
\newcommand{\en}{\end{eqnarray}}

\def\bea{\begin{eqnarray}}
\def\eea{\end{eqnarray}}

\usepackage{amsmath}
\usepackage{amssymb}
\usepackage{graphicx}
\usepackage{bm}
\usepackage{epstopdf}
\usepackage{slashed}
\usepackage{color}

\usepackage{graphics}

\usepackage[unicode=true,
 bookmarks=true,bookmarksnumbered=false,bookmarksopen=false,
 breaklinks=false,pdfborder={0 0 1},backref=false,colorlinks=false]
 {hyperref}

\begin{document}

\title{New point-like sources and a conducting surface in Maxwell-Chern-Simons electrodynamics}

\author{L.H.C. Borges$^{1}$}
\email{luizhenriqueunifei@yahoo.com.br}

\author{F.E. Barone}
\email{frederico.barone@gmail.com}

\author{C.C.H. Ribeiro$^{2}$}
\email{caiocesarribeiro@ifsc.usp.br}

\author{H.L. Oliveira$^{1}$}
\email{helderluiz10@gmail.com}

\author{R.L. Fernandez$^{3}$}
\email{rlipparelli@if.ufrj.br}

\author{F.A. Barone$^{4}$}
\email{fbarone@unifei.edu.br}

\affiliation{$^{1}$UNESP - Campus de Guaratinguet\'a - DFQ, Avenida Dr. Ariberto Pereira da Cunha 333, 
CEP 12516-410, Guaratinguet\'a, SP, Brazil}

\affiliation{$^{2}$Instituto de F\'\i sica de S\~ao Carlos, Universidade de S\~ao Paulo, S\~ao Carlos, S\~ao Paulo 15980-900, Brazil}

\affiliation{$^{3}$Instituto de F\'\i sica - Universidade Federal do Rio de Janeiro - Av. Athos da Silveira Ramos, 149 
Centro de Tecnologia - bloco A - Cidade Universitária - Rio de Janeiro - RJ - CP: 68528 - CEP: 21941-972}

\affiliation{$^{4}$IFQ - Universidade Federal de Itajub\'a, Av. BPS 1303, Pinheirinho, Caixa Postal 50, 37500-903, Itajub\'a, MG, Brazil}

\begin{abstract}
We investigate some aspects of the Maxwell-Chern-Simons electrodynamics focusing on physical effects produced by the presence of stationary sources and a perfectly conducting plate (mirror). Specifically, in addition to point charges, we propose two new types of point-like sources called topological source and Dirac point, and we also consider physical effects in various configurations that involve them. We show that the Dirac point is the source of the vortex field configurations. The propagator of the gauge field due to the presence of a conducting plate and the interaction forces between the plate and  point-like sources are computed. It is shown that the image method is valid for the point-like charges as well as for Dirac points. For the topological source we show that the image method is not valid and the symmetry of spatial refection on the mirror is broken.  In all setups considered, it is shown that the topological source leads to the emergence of torques.  
\end{abstract}

\maketitle

\section{Introduction}
\label{introduction}

Planar models in Quantum Field Theory have several interesting features, both theoretical and experimental.  We can mention, for instance, the change in the fermions behavior in comparison with the standard classical and quantum electrodynamics \cite{Dunne}. One of the most important class of models of this kind is the so called Maxwell-Chern-Simons electrodynamics \cite{MCS1}, or Abelian topological massive gauge theory \cite{MCS2}, which is relevant because it is simultaneously massive and gauge invariant. Theoretical aspects of Maxwell-Chern-Simons electrodynamics have been investigated in Casimir effect  \cite{Casimir1,Casimir2,Casimir3,Casimir4,Casimir5,Casimir6}, quantum dissipation of harmonic systems \cite{Dissipation}, quantum electrodynamics (QED$_{3}$) \cite{QED1,QED2,QED3,QED4,QED5}, dynamical mass generation  \cite{DMG1,DMG2}, condensed matter physics (see, for instance, Ref. \cite{CMP} and references therein), description of graphene properties \cite{Nature,PRDGraphene,PRB97Graphene,EPJBGrphene,APGraphene,PRB99Graphene}, noncommutativity \cite{NC1,NC2,NC3,NC4}, strings theory \cite{string}, dynamics of vortices \cite{V1,V2}, and with a planar boundary \cite{Planar1,Planar2,Planar3,Planar4,Planar5}, to mention just a few. In fact, there is a vast literature concerning this model.

There is also a generalization of the Chern-Simos electrodynamics in $3+1$ dimensions, the so called Carroll-Field-Jackiw model \cite{CFJ}, which exhibits Lorentz symmetry breaking and whose corresponding electrostatics and magnetostatics has been studied thoroughly in reference \cite{PRD2004}, as well as the Casimir Effect, in references \cite{PRD81!025015,PRD95!036011}. Another coupling involving the dual gauge field strength tensor in $3+1$ dimensions is the so called axion $\theta$-electrodynamics, which can be used to describe insulators with boundaries \cite{axion1,axion2,axion3,EPL113!60005}. 

In the context of Casimir Effect, in $3+1$ dimensions, Chern-Simons surfaces can also be used to obtain Casimir repulsion setups with planar symmetry \cite{PRB97!165432}. In higher dimensions, the Casimir force has been studied in Randall-Sundrum models \cite{PRD76!015008}, which can be interpreted as a kind of ground state for Chern-Simons gravity \cite{CQC33!235004}.  

Regarding the Maxwell-Chern-Simons electrodynamics, there are two interesting questions no yet explored in the literature, to the best of the authors knowledge. The first one concerns the physical phenomena produced by the presence of point-like field sources. The second one concerns the modifications which the propagator undergoes due to the presence of a single perfectly conducting plate (mirror), as well as the interaction between mirrors and point-like sources.

In this paper we delve deeper in these topics by searching for physical effects produced by the presence of stationary  point-like sources and a perfectly conducting plate. Specifically, in section \ref{fontes} we study the interactions between  pairs of external sources mediated by the Maxwell-Chern-Simons field. In addition to the point charges we analyze configurations involving two kinds of sources not considered previously in the literature, as far as the authors know. We call these sources Dirac points and topological sources. It is shown that the topological source generalizes the Dirac point. In all setups  considered, we show that the topological source leads to the emergence of torques. In section \ref{MCSEMF} we compute the field configurations generated by the point-like sources  and show that the Dirac source leads to vortex configurations for the gauge field. In section \ref{MCSplaca} we consider the Maxwell-Chern-Simons field in the presence of a conducting plate and obtain the corresponding propagator to study the interaction between the conductor and the sources. We also compare the interaction forces with the ones obtained in the free theory (theory without the plate) and we verify that the image method is valid for the point-like charges as well as for Dirac points. We show that the image method is not valid for the interaction between the conducting plate and the topological source. As consequence of this feature, we have that the symmetry of spatial reflection on the mirror is broken. We also show the emergence of a torque acting on a setup where the distance between the topological source and the plate is kept fixed. Finally, section \ref{conclusoes} is devoted to our final comments.

In this paper we work in a $2+1$-dimensional Minkowski space-time  with metric $\eta^{\mu\nu}=(1,-1,-1)$. The Levi-Civita tensor is denoted by $\epsilon^{\mu\nu\lambda}$ with $\epsilon^{012}=1$.

\section{Point-like sources for the Maxwell-Chern-Simons field}
\label{fontes}

The Maxwell-Chern-Simons Lagrangian, defined in a $2+1$ spacetime, is given by
\begin{eqnarray}
\label{Lagrangian}
{\cal L}= -\frac{1}{4}F_{\mu\nu}F^{\mu\nu}-\frac{1}{2\gamma}\left(\partial_{\mu}A^{\mu}\right)^{2}+
\frac{1}{2}\ m\epsilon^{\mu\nu\lambda}A_{\mu}\partial_{\nu}A_{\lambda}-J^{\mu}A_{\mu} \ ,
\end{eqnarray}
where $A^{\mu}$ is the gauge field, $F^{\mu\nu}=\partial^{\mu}A^{\nu}-\partial^{\nu}A^{\mu}$ is the field strength, $J^{\mu}$ is the external source, $\gamma$ is a gauge parameter and $m$ is a mass parameter.

The external source must have null divergence in order to assure gauge invariance of the last term in (\ref{Lagrangian}), namely, $\partial_{\mu}J^{\mu}=0$

The model (\ref{Lagrangian}) is equivalent to
\begin{eqnarray}
\label{LagrnagianOP}
{\cal L}\rightarrow \frac{1}{2}A^{\mu}\left[\eta_{\mu\nu}\partial_{\lambda}\partial^{\lambda}-\left(1-\frac{1}{\gamma}\right)\partial_{\mu}\partial_{\nu}-m\epsilon_{\mu\nu\lambda}\partial^{\lambda}\right]A^{\nu}-J^{\mu}A_{\mu} \ .
\end{eqnarray}
Using the Feynman gauge, where $\gamma=1$, the corresponding propagator reads \cite{RBEF2011}
\begin{eqnarray}
\label{propagator}
D^{\mu\nu}\left(x,y\right)=-\int\frac{d^{3}p}{(2\pi)^{3}}\frac{1}{p^{2}-m^{2}}
\left(\eta^{\mu\nu}-m^{2}\frac{p^{\mu}p^{\nu}}{p^{4}}+i\frac{m}{p^{2}}
\epsilon^{\mu\nu\lambda}p_{\lambda}\right)e^{-ip\cdot(x-y)} \ ,
\end{eqnarray}
in the sense that
\begin{eqnarray}
\label{propO}
\left[\eta_{\mu\nu}\partial_{\lambda}\partial^{\lambda}-m\epsilon_{\mu\nu\lambda}\partial^{\lambda}\right]D^{\nu\sigma}\left(x,y\right)=\eta_{\mu}^{\ \sigma}\delta^{3}\left(x-y\right) \ .
\end{eqnarray}

As discussed in references \cite{Zee,BaroneHidalgo1,BaroneHidalgo2}, the contribution due to the sources to the ground state energy of the system is given by
\begin{equation}
\label{zxc1}
E=\frac{1}{2T}\int\int d^{3}x\ d^{3}y J^{\mu}(x)D_{\mu\nu}(x,y)J^{\nu}(y)\ ,
\end{equation}
where $T$ is the time variable.

In the first setup, we consider the field sources given by
\begin{eqnarray}
\label{corre1Em}
J^{CC}_{\mu}({\bf x})=\sigma_{1}\eta_{\ \mu}^{0}\delta^{2}\left({\bf x}-{\bf a}_ {1}\right)+\sigma_{2}\eta_{\ \mu}^{0}\delta^{2}\left({\bf x}-{\bf a}_ {2}\right) \ ,
\end{eqnarray}
where we have two spatial Dirac delta functions, concentrated at the positions ${\bf a}_{1}$ and ${\bf a}_{2}$. The parameters $\sigma_{1}$ and $\sigma_{2}$ are the coupling constants among the field and the delta functions and can be interpreted as electric charges. Henceforth the superscript $CC$ means that we have the interaction  between two point charges. 

Substituting (\ref{propagator}) and (\ref{corre1Em}) in (\ref{zxc1}), discarding the self-interacting contributions (the interactions of a given point-charge with itself), performing the integrals in the following order,  $d^{2}{\bf x}$, $d^{2}{\bf y}$, $dx^{0}$, $dp^{0}$ and $dy^{0}$, using the Fourier representation for the Dirac delta function, $\delta(p^{0})=\int dx/(2\pi)\exp(-ipx^{0})$, and identifying the time interval as $T=\int dy^{0}$, we obtain
\begin{eqnarray}
\label{Ener2EM}
E^{CC}=\sigma_{1}\sigma_{2}\int\frac{d^{2}{\bf p}}{(2\pi)^{2}}\frac{\exp(i{\bf p}\cdot{\bf a})}{{\bf p}^2+m^2} \ ,
\end{eqnarray}
where we defined ${\bf{a}}={\bf {a}}_{1}-{\bf {a}}_{2}$, which is the distance between the two electric charges.

Using the fact that \cite{BaroneHidalgo1}
\begin{eqnarray}
\label{int4EM}
\int\frac{d^{2}{\bf p}}{(2\pi)^{2}}\frac{\exp(i{\bf p}\cdot{\bf a})}
{{\bf p}^2+m^2}=\frac{1}{2\pi}K_{0}(ma) \ ,
\end{eqnarray}
where $a=\mid\bf{a}\mid$, and $K$ stands for the K-Bessel function \cite{Arfken} , we can write
\begin{eqnarray}
\label{Ener3EM}
E^{CC}=\frac{\sigma_{1}\sigma_{2}}{2\pi}K_{0}(ma) \ .
\end{eqnarray}

Therefore, the interacting force between two charges is given by 
\begin{eqnarray}
\label{For1EM}
{\bf F}^{CC}=-\frac{\partial E^{CC}}{\partial{\bf a}}=\frac{m\sigma_{1}\sigma_{2}}{2\pi}K_{1}(ma){\hat a} \ .
\end{eqnarray}
which is an  usual result in theories of massive fields.

Let us see if we can find other kinds of interactions with not so trivial sources. For this task we propose a second kind of point-like external source to the Maxwell-Chern-Simons field. We shall call it topological source and we start by considering a system compose by two topological sources placed at the positions ${\bf{a}}_{1}$ and ${\bf{a}}_{2}$, as follows
\begin{eqnarray}
\label{sourceet2}
J^{TT}_{\mu}\left(x\right)=\epsilon_{\mu}^{\ \alpha\beta}V_{\alpha}
\partial_{\beta}\delta^{2}\left({\bf x}-{\bf a}_{1}\right) 
+\epsilon_{\mu}^{\ \alpha\beta}U_{\alpha}\partial_{\beta}
\delta^{2}\left({\bf x}-{\bf a}_{2}\right)\ ,
\end{eqnarray}
where the superscript $TT$ means that we have the interaction between two topological sources. 

In expression (\ref{sourceet2}), $V_{\alpha}=(V^{0},{\bf V})$ and $U_{\alpha}=(U^{0},{\bf U})$ are two constant Minkowski $3$-pseudo-vectors. Each term of Eq. (\ref{sourceet2}) can be obtained from the point-like source proposed in reference \cite{BaroneBaroneHelayel} for the Kalb-Ramond field, by dimensional reduction.

We do not need to impose, in an {\it ad hoc} way, that the source (\ref{sourceet2}) satisfies the continuity equation. From the anti-symmetry of the differential operator $\epsilon_{\mu}^{\ \alpha\beta}\partial_{\beta}\partial^{\mu}$, we can ensure that $\partial^{\mu}J^{TT}_{\mu}\left(x\right)=0$. So the source (\ref{sourceet2}) leads to an intriscic conserved quantity $\int d^{2}{\bf r}J^{TT}_{0}$. In addition, if we write the action term which couples the source (\ref{sourceet2}) to the gauge field, $J^{TT}_{\mu}A^{\mu}$, in a curved space-time, with the substitution $\eta^{\mu\nu}\to g^{\mu\nu}$, we can see that this term does not couples to the gravitational field, similarly to what happens to the Chern-Simons term \cite{Zee}. This is why we named (\ref{sourceet2}) as topological source.

Here, some points are in order. A partity transformation (for polar vectors) must have determinant equal do $-1$, independent of the space dimensionality. So, in 3 dimensions, a parity transformation can be a complete spatial inversion (the most common definition) or even just the inversion of a single cartesian variable. In 2 dimensions, a parity transformation must be the inversion of just a single spatial cartesian variable. The behavior of a pseudo-vector under a parity transformation is the contrary to the one exhibited by a true vector.

It is interesting to notice that for the very specific case where $V_{0}=0$ and $U_{0}=0$, each term in the source (\ref{sourceet2}) corresponds to the charge distribution of an electric dipole \cite{BaroneHidalgo2}, the first one defined by ${\bf d}_{V}=(V^{2},-V^{1})$ and the second one, by ${\bf d}_{U}=(U^{2},-U^{1})$. Taking into account that ${\bf V}$ and ${\bf U}$ are axial vectors and their behavoiurs under a parity transformation, one can show that ${\bf d}_{V}$ and ${\bf d}_{U}$ are polar vectors. If we think on the proposed (2+1)-dimensional model as embedded in a (3+1) dimensional space-time, we could write, for instance,
\begin{eqnarray}
\label{defd}
{\bf d}_{V}=(V^{2},-V^{1},0)={\bf V}\times{\hat z}\cr\cr
{\bf d}_{U}=(U^{2},-U^{1},0)={\bf U}\times{\hat z}\ .
\end{eqnarray}
what evinces that ${\bf d}_{V}$ is a polar vector, once ${\bf V}$ is an axial vector. 

Substituting the source (\ref{sourceet2}) in expression (\ref{zxc1}), discarding the self-interacting contributions and proceeding as we have done previously, we obtain   
\begin{eqnarray}
\label{EnerTEM}
E^{TT}&=&\Bigg[V^{0}U^{0}{\bf\nabla}^{2}_{{\bf a}}-[({\bf{V}}\times{\hat z})\cdot{\bf\nabla}_{\bf a}][({\bf{U}}\times{\hat z})\cdot{\bf\nabla}_{\bf a}]\cr\cr
&\ &+m[(U^{0}{\bf V}-V^{0}{\bf U})\times{\bf\nabla}_{\bf a}]\cdot{\hat z}\Bigg]
\int\frac{d^{2}{\bf p}}{(2\pi)^{2}}\frac{\exp(i{\bf p}\cdot{\bf a})}{{\bf p}^2+m^2}
\end{eqnarray}
where $i,j=1,2$ are spatial indexes, ${\bf\nabla}_{{\bf a}}^{i}=\partial/\partial a^{i}$ and we defined the differential operator
\begin{eqnarray}
\label{exchange}
{\bf\nabla}_{{\bf a}}=\left(\frac{\partial}{\partial a^{1}},\frac{\partial}{\partial a^{2}}\right) \ .
\end{eqnarray}

Substituting the result (\ref{int4EM}) in the energy (\ref{EnerTEM}) and carrying out the calculations, we find
\begin{eqnarray}
\label{EnerTEM2}
E^{TT}&=&\frac{m^{2}}{2\pi}\Bigg[ V^{0}U^{0}K_{0}(m a)+{\bf{V}}\cdot{\bf{U}}\frac{K_{1}(ma)}{ma}\nonumber\\
& &-\frac{({\bf V}\times{\hat z})\cdot{\bf a}}{a}\frac{({\bf U}\times{\hat z})\cdot{\bf a}}{a}K_{2}(ma)\nonumber\\
& &+\left[{\bf a}\times\left(U^{0}{\bf V}-V^{0}{\bf U}\right)\right]\cdot{\hat z}\frac{K_{1}(ma)}{a}\Bigg]\ .
\end{eqnarray}
where we used the fact that
\begin{eqnarray}
K_{0}(ma)+2\frac{K_{1}(ma)}{ma}=K_{2}(ma)\ .
\end{eqnarray}

The interaction force between two topological field sources is then:
\begin{eqnarray}
\label{fortptp} 
{\bf{F}}^{TT}&=&-\frac{\partial E^{TT}}{\partial{\bf a}}=
\frac{m^{2}}{2\pi}\Bigg[V^{0}U^{2}mK_{1}(ma)+{\bf V}\cdot{\bf U}\frac{K_{2}(ma)}{a}\cr\cr
&\ &-\frac{({\bf V}\times{\hat z})\cdot{\bf a}}{a}\frac{({\bf U}\times{\hat z})\cdot{\bf a}}{a}\left(mK_{1}(ma)+\frac{4K_{2}(ma)}{a}\right)\cr\cr
&\ &+\left({\bf V}\times{\hat z}\frac{({\bf U}\times{\hat z})\cdot{\bf a}}{a}+\frac{({\bf V}\times{\hat z})\cdot{\bf a}}{a}{\bf U}\times{\hat z}\right)\frac{K_{2}(ma)}{a}\cr\cr
&\ &+\left[{\bf a}\times\left(U^{0}{\bf V}-V^{0}{\bf U}\right)\right]\cdot{\hat z}\frac{mK_{2}(ma)}{a}\cr\cr
&\ &-\left(U^{0}{\bf V}-V^{0}{\bf U}\right)\times{\hat z}\frac{K_{1}\left(m a\right)}{a}\Bigg]\ .
\end{eqnarray}
The force (\ref{fortptp}) exhibits a strong anisotropic behavior and decreases with Bessel functions when $a$ increases.

In the case where $V^{0}=U^{0}=0$, we can use the expressions (\ref{defd}) to show that the energy (\ref{EnerTEM2}) becomes exactly the same one found for the interaction between two typical electric dipoles in a $2+1$ dimensional theory with a massive vector field \cite{BaroneHidalgo2}.

It is important to mention that the topological source also produces effects in the standard Maxwell electrodynamics in 2+1 dimensions. In order to verify this fact we must take the limit $m\rightarrow 0$ in Eq. (\ref{EnerTEM2}). In this case, all terms which depend on $V^{0}$ or $U^{0}$ vanishes, and we are taken to the same interaction obtained for two electric dipoles (\ref{defd}), namely 
\begin{eqnarray}
\label{fortptpm}
E^{TT}\left(m=0\right)=E^{TT}_{M}&=&
\frac{1}{2\pi a^{2}}\Biggl[\left({\bf{V}}\cdot{\bf{U}}\right)-2\frac{({\bf V}\times{\hat z})\cdot{\bf a}}{a}\frac{({\bf U}\times{\hat z})\cdot{\bf a}}{a}
\Biggr] \ .
\end{eqnarray}
The subscript $M$ in (\ref{fortptpm}) means that we have the quantities calculated for the Maxwell theory in three dimensions. 

Notice that just the spatial parts of $U^{\mu}$ and $V^{\mu}$ are relevant for the energy (\ref{fortptpm}), where we have the massless case. Using definition (\ref{defd})and the fact that ${\bf V}\cdot{\bf U}={\bf d}_{V}\cdot{\bf d}_{U}$, it can be shown that (\ref{fortptpm})  has the same behavior as the one found with two electric dipoles. The force is still anisotropic and decreases  with  distance $a$.

\begin{figure}[!h]
\centering
\includegraphics[scale=0.2]{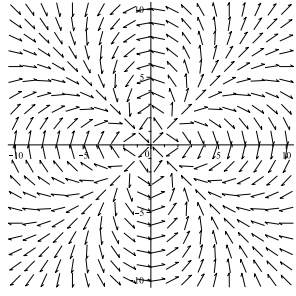}
\caption{Plot for the force obtained from (\ref{fortptpm}), with ${\bf V}=V_x{\hat x}$ and ${\bf U}=U_y{\hat y}$. Vertical axis is $a_{y}$ and horizontal axis is $a_{x}$.}
\label{Eq16a}
\end{figure}

It is interesting to analyze the force obtained from the energy (\ref{fortptpm}) in terms of $U^{\mu}$ and $V^{\mu}$ (and not in terms of the respective electric dipoles). For this task we take the specific and simple case where ${\bf V}=V_x{\hat x}$ and ${\bf U}=U_y{\hat y}$ and plot in figure (\ref{Eq16a}) the force-lines obtained from (\ref{fortptpm}) multiplied by $a^{3}/(V_{x}U_{y})$. The vertical axis is the component $a_{y}$ and the horizontal one, $a_{x}$.

\begin{figure}[!h]
\centering
\includegraphics[scale=0.4]{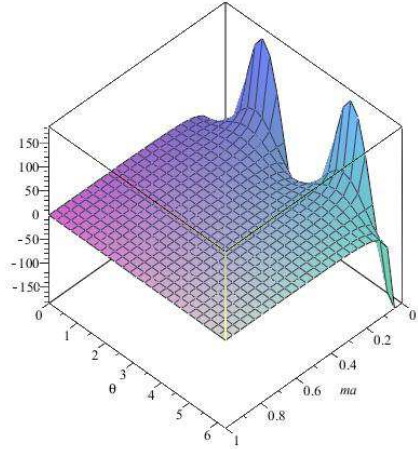}
\caption{Plot for the torque in (\ref{TTT}), ${\tau}^{TT}/(m^{2}V_{x}U_{y})$.}
\label{Eq18a}
\end{figure}

The asymmetry in the interaction (\ref{fortptpm}) (and also in (\ref{EnerTEM2})) brings on torques on the topological sources. Taking the same configuration considered previously, with ${\bf V}=V_{x}{\hat x}$, ${\bf U}=U_{y}{\hat y}$, and for simplicity taking $V^{0}=U^{0}=0$, and next writing ${\bf a}=a[\cos(\theta){\hat x}+\sin(\theta){\hat y}]$, with $\theta$ standing for the usual azimuthal angle in polar coordinates, the energy (\ref{EnerTEM2}) becomes 
\begin{equation}
E^{TT}(V^{0}=U^{0}=0,\theta)=\frac{m^{2}}{2\pi}V_{x}U_{y}\cos(\theta)\sin(\theta)K_{2}(m a)\ ,
\end{equation}
and we have a torque on the whole system 
\begin{eqnarray}
\label{TTT}
{\tau}^{TT}&=&-\frac{\partial}{\partial\theta}E^{TT}(V^{0}=U^{0}=0,\theta)
=\frac{m^{2}}{2\pi}V_{x}U_{y}(\sin^{2}(\theta)-\cos^{2}(\theta))K_{2}(m a)\ .
\end{eqnarray}
which is plotted in fig(\ref{Eq18a})

Just for completeness, we point out that in the standard Maxwell electrodynamics (in 2+1 dimensions), we also have a torque on the corresponding system. This fact can be verified by taking the limit $m\rightarrow 0$ in Eq. (\ref{TTT}), 
\begin{eqnarray}
\label{TTTM}
{\tau}^{TT}_{M}=\frac{V_{x}U_{y}}{\pi{{a}}^{2}}(\sin^{2}(\theta)-\cos^{2}(\theta)) \ .
\end{eqnarray}

Now we define another non-trivial external source, as follows 
\begin{eqnarray}
\label{dpcharge}
J^{\mu}_{(D)}\left({\bf x}\right)=-2\pi i\Phi\int\frac{d^{3}p}{\left(2\pi\right)^{3}} \ 
\delta\left(p^{0}\right)\epsilon^{0\mu\alpha}p_{\alpha} \ 
e^{-i p\cdot x}e^{-i {\bf{p}}\cdot{{\bf a}_{1}}} \ ,
\end{eqnarray}
where $\Phi$ is a constant with dimension of magnetic flux, and ${\bf a}_{1}$ is a spatial-vector.

Eq. (\ref{dpcharge}) is obtained by dimensional reduction of the source which describes a Dirac string in $3+1$ dimensions \cite{DiracString1,DiracString2,DiracString3,DiracString4,DiracString5} lying along the axis $(x^{1},x^{2},x^{3})=({\bf a}_{1},x^{3})$, parallel to the $3$-axis, with internal magnetic flux $\Phi$. So, from now on, we shall call the source (\ref{dpcharge}) as Dirac point.

It is possible to show that the topological source is the generalization of the Dirac point. For this task we take  the first term on the right hand side of Eq. (\ref{sourceet2}), for instance, in the specific case where $V^{0}=-\Phi$ and ${\bf V}=0$ ($V^{\mu}=-\Phi\eta^{\mu0}$). In this situation, this term becomes exactly the right hand side of Eq. (\ref{dpcharge}). 

In this way, taking ${\bf U}={\bf V}=0$ in Eq. (\ref{EnerTEM2}), we obtain the interaction energy between two Dirac-points, where the first one is located at the position ${\bf{a}}_{1}$, with magnetic flux $V^{0}=-\Phi_{1}$ and the second one placed at the position ${\bf{a}}_{2}$ with magnetic flux $U^{0}=-\Phi_{2}$. So that, the energy (\ref{EnerTEM2}) becomes
\begin{eqnarray}
\label{Ener5EM}
E^{TT}\left({\bf U}={\bf V}=0,V^{0}=-\Phi_{1}, U^{0}=-\Phi_{2} \right)=E^{DD}=\frac{m^{2}\Phi_{1}\Phi_{2}}{2\pi}K_{0}(m a) \ ,
\end{eqnarray}
where the superscript $DD$ means that we have a system composed by two Dirac points.

The interaction energy in Eq. (\ref{Ener5EM})  is an effect due solely to Maxwell-Chern-Simons electrodynamics,  because if we take the limit $m\rightarrow 0$ there is no interaction energy. On the contrary to the energy (\ref{EnerTEM2}), which is non-vanishing for $m=0$, as we can see in (\ref{fortptpm}).

From the Eq. (\ref{fortptp}), the interaction force between two Dirac points is given by
\begin{eqnarray}
\label{For2EM}
{\bf{F}}^{TT}\left({\bf U}={\bf V}=0,V^{0}=-\Phi_{1}, U^{0}=-\Phi_{2} \right)={\bf{F}}^{DD}=\frac{m^{3}\Phi_{1}\Phi_{2}}{2\pi}K_{1}(m a){\hat{a}}  \ .
\end{eqnarray}

It is quite interesting to notice the similarity between expressions (\ref{Ener5EM}) and (\ref{Ener3EM}). Identifying $\sigma_{1}=m\Phi_{1}$ and $\sigma_{2}=m\Phi_{2}$ in (\ref{Ener5EM}), we are taken to Eq. (\ref{Ener3EM}) and the Dirac points behave like two point-like charges. In the next section we show that the Dirac point (\ref{dpcharge}) is the field source for a vortex solution for the gauge field.

It is known in the literature that the presence of a monopole with an axion-like term in (3+1) dimensions can lead to an effective electric charge seen by an observer far away from the monopole \cite{PRL!58!1799}. In some sence, Eq. (\ref{For2EM}) resembles this result in the (2+1)-dimensional model considered in this paper. In addition, if we define the vector
\begin{eqnarray}
\label{defvecK}
{\bf K}\left({\bf x}\right)=-2\pi i\Phi\int\frac{d^{3}p}{\left(2\pi\right)^{3}} \ 
\delta\left(p^{0}\right){\bf p} e^{-i p\cdot x}e^{-i {\bf{p}}\cdot{\bf {a}_{1}}}\cr\cr
=\nabla_{{\bf a}_{1}}\left[2\pi\Phi\int\frac{d^{3}p}{\left(2\pi\right)^{3}}\delta\left(p^{0}\right)e^{-i p\cdot x}e^{-i {\bf{p}}\cdot{\bf {a}_{1}}}\right] \ ,
\end{eqnarray}
we can rewrite the source (\ref{dpcharge}) in the compact form $J^{\mu}_{(D)}\left({\bf x}\right)={\hat z}\times{\bf K}\left({\bf x}\right)$. Noticing that the vector (\ref{defvecK}) resembles the vortex Hall current which produces magnetic monopole-type fields in topological insulators, we could conjecture if the Dirac source would not be related to the vortex Hall current. This subject deserves more investigations and would render the source (\ref{dpcharge}) a candidate to study vortex Hall currents.

Just for completeness, we consider the interaction between a topological source and a point-like charge, with the source
\begin{eqnarray}
\label{corre1Emmm}
J^{CT}_{\mu}({\bf x})=\sigma\eta_{\ \mu}^{0}\delta^{2}\left({\bf x}-{\bf a}_ {1}\right)+
\epsilon_{\mu}^{\ \alpha\beta}V_{\alpha}\partial_{\beta}
\delta^{2}\left({\bf x}-{\bf a}_{2}\right) \ ,
\end{eqnarray}
where the point-like charge is placed at position ${\bf{a}}_{1}$ and the topological
field source is placed at position ${\bf{a}}_{2}$. The superscript $CT$ means that we have the interaction between a point-like charge and a topological source. 

The interaction energy is then given by,
\begin{eqnarray}
\label{dpplc}
E^{CT}&=&-\sigma\left({\bf V}\times{\bf\nabla}_{{\bf a}}
+m V^{0}\right)\int\frac{d^{2}{\bf p}}
{(2\pi)^{2}}\frac{\exp(i{\bf p}\cdot{\bf a})}{{\bf p}^2+m^2}\nonumber\\
&=&-\frac{m\sigma}{2\pi}\left[\frac{K_{1}\left(m a\right)}{a}
({\bf V}\times{\hat z})\cdot{\bf a}+V^{0}K_{0}\left(m a\right)\right] \ ,
\end{eqnarray}
with the corresponding interaction force
\begin{eqnarray}
\label{fortpch} 
{\bf{F}}^{CT}&=&-{\bf\nabla}_{{\bf a}}E^{CT}\nonumber\\
&=&\frac{m^{2}\sigma}{2\pi}\Bigl\{{\bf V}\times{\hat z}\frac{K_{1}\left(m a\right)}{ma}\nonumber\\
&\ &-\Bigl[\frac{K_{2}\left(m a\right)}{a}({\bf V}\times{\hat z})\cdot{\bf a}+V^{0}K_{1}\left(m a\right)\Bigr]{\hat{a}}\Bigr\} \ .
\end{eqnarray}

Notice that (\ref{dpplc}) is the interaction energy between a charge $\sigma$ with an effective charge $-mV^{0}$ added by the interaction energy between a charge $\sigma$ and an effective electric dipole ${\bf V}\times{\hat z}$.

For the Maxwell electrodynamics (limit $m\rightarrow 0$) the interaction energy (\ref{dpplc}) reads
\begin{eqnarray}
\label{fortpchM}
E^{CT}(m=0)=E^{CT}_{M}=\frac{\sigma}{2\pi}\frac{1}{a^{2}}({\bf V}\times{\hat z})\cdot{\bf a}
\end{eqnarray}

\begin{figure}[!h]
\centering
\includegraphics[scale=0.2]{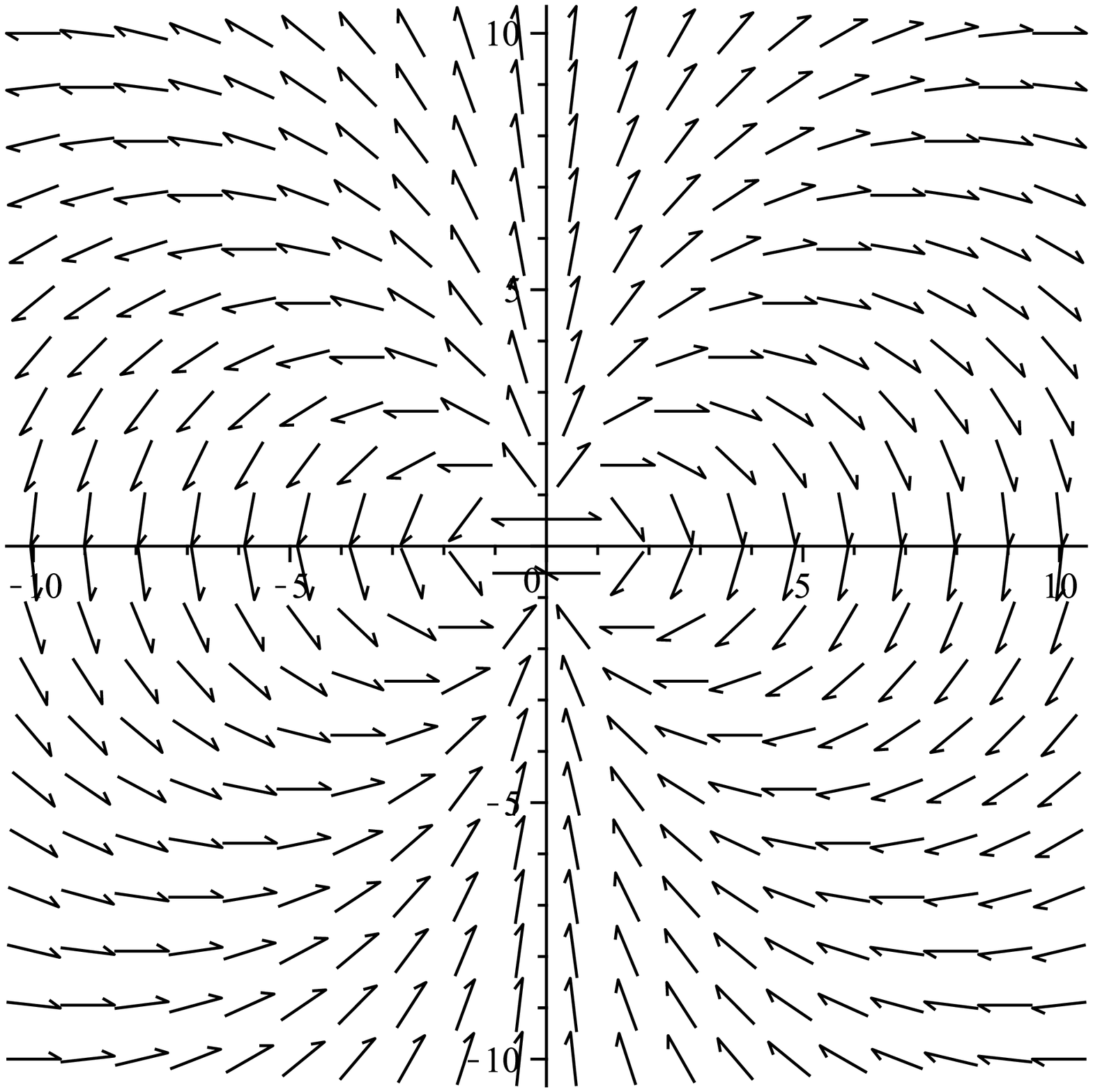}
\caption{Plot for the normalized force lines obtained from (\ref{fortpchM}) with ${\bf V}=V_{x}{\hat x}$ and ${\bf a}$ centered at the origin.}
\label{Eq26a}
\end{figure}

In order to see the anisotropic features of the force produced by the energy (\ref{fortpchM}), with respecto to ${\bf V}$, we restrict to the specific case where ${\bf V}=V_{x}{\hat x}$ and take ${\bf a}$ centered at the origin to plot the normalized force lines corresponding to (\ref{fortpchM}) in figure (\ref{Eq26a}).

By fixing the distance $a$ between the point-like charge and the topological source, we can show that the energy (\ref{dpplc}) leads to a torque on the setup with respect to the vector ${\bf a}$. For simplicity, we take the specific situation where $V^{0}=0$, ${\bf V}=V_{x}{\hat x}$ and ${\bf a}=a[\cos(\theta){\hat x}+\sin(\theta){\hat y}]$, obtaining
\begin{equation}
E^{CT}(V^{0}=0,\theta)=\frac{m\sigma}{2\pi}V_{x}\sin(\theta)K_{1}\left(m a\right)\ ,
\end{equation}
\begin{eqnarray}
\label{torct}
\tau^{CT}&=&-\frac{\partial}{\partial\theta}E^{CT}(V^{0}=0,\theta)=-\frac{m\sigma}{2\pi}V_{x}\cos(\theta)K_{1}\left(m a\right) \ .
\end{eqnarray}

\begin{figure}[!h]
\centering
\includegraphics[scale=0.3]{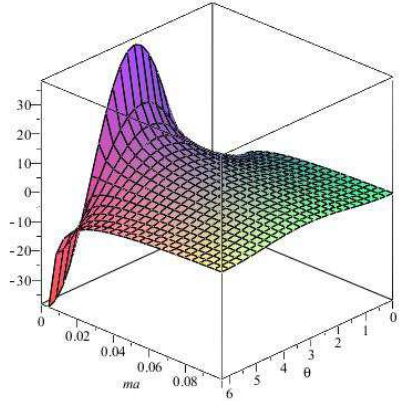}
\caption{Plot for the torque $\tau^{CT}/(m\sigma V_{x})$ in Eq.(\ref{torct}).}
\label{Eq28a}
\end{figure}

Taking the limit $m\rightarrow 0$ in the expression (\ref{torct}), we have, once again, a torque in Maxwell electrodynamics due to the presence of the topological source,
\begin{eqnarray}
\label{torctM}
\tau^{CT}_{M}=-\frac{\sigma V_{x}}{2\pi a}\cos(\theta) \ .
\end{eqnarray}

When ${\bf V}=0$, identifying the flux $V^{0}=-\Phi$, Eq. (\ref{dpplc}) becomes the interaction energy between a point-like charge placed at position ${\bf{a}}_{1}$ and a Dirac point placed at position ${\bf{a}}_{2}$, as follows
\begin{equation}
\label{Ener101EM}
E^{CT}\left(V^{0}=-\Phi,{\bf V}=0 \right)=E^{CD}=\frac{m\sigma\Phi}{2\pi}K_{0}\left(ma\right) \ ,
\end{equation}
where the superscript $CD$ means that we have the interaction between a point-like charge and a Dirac point. 
This interaction energy has no counterpart in Maxwell electrodynamics.

From  Eq. (\ref{fortpch}), the interaction force between the charge and the Dirac point reads
\begin{eqnarray}
\label{For111EM}
{\bf{F}}^{CT}\left(V^{0}=-\Phi,{\bf V}=0 \right)={\bf{F}}^{CD}=\frac{m^{2}\sigma\Phi}{2\pi}K_{1}(ma){\hat{a}} \ .
\end{eqnarray}

Once again, it is remarkable the similarity between expression (\ref{Ener101EM}) and (\ref{Ener3EM}). If we identify $\sigma_{eff}=m\Phi$ in Eq. (\ref{Ener101EM}), we can see that the Dirac point behaves like a point-charge in the interaction with another point-like charge.

\section{Electromagnetic field}
\label{MCSEMF}

In this section we calculate the electromagnetic field configurations produced by all the point-like sources considered
in previous sections. For this task, we choose a coordinate system where the corresponding external source 
is placed at the origin and compute the electromagnetic field evaluated at an arbitrary point 
${\bf{r}}=\left(r^{1},r^{2}\right)$.

The field configuration can be calculated with the aid of the propagator ({\ref{propagator}}), as follows
\begin{equation}
\label{AMU}
A^{\mu}\left(r\right)=\int d^{3}y D^{\mu\nu}\left(r,y\right)J_{\nu}\left(y\right) \ ,
\end{equation}
where $A^{\mu}=\left(A^{0},A^{1},A^{2}\right)$. 

Let us start by considering a topological source concentrated  at origin, 
\begin{equation}
\label{topori}
J^{T}_{\nu}({y})=\epsilon_{\nu}^{\ \alpha\beta}V_{\alpha}\partial_{\beta}
\delta^{2}\left({\bf y}\right) \ .
\end{equation}
Substituting (\ref{propagator}) and (\ref{topori}) in  expression (\ref{AMU}), and then performing
some manipulations similar to the ones employed in the previous section, we obtain
\begin{eqnarray}
\label{AOE}
A^{0}\left({\bf{r}}\right)&=&
-\left[\left({\bf V}\times{\bf\nabla}_{{\bf r}}\right)\cdot{\hat z}+m V^{0}\right]\int\frac{d^{2}{\bf p}}
{(2\pi)^{2}}\frac{\exp(i{\bf p}\cdot{\bf r})}{{\bf p}^2+m^2}\nonumber\\
&=&-\frac{m}{2\pi}\left[V^{0}K_{0}\left(mr\right)+\frac{K_{1}\left(mr\right)}
{r}({\bf V}\times{\hat z})\cdot{\bf r}\right] \ .
\end{eqnarray}

For the spatial components ${\bf A}\left({\bf{r}}\right)$ we have
\begin{eqnarray}
\label{AKE}
{\bf A}\left({\bf{r}}\right)&=&V^{0}{\hat z}\times{\bf\nabla}_{{\bf r}}
\int\frac{d^{2}{\bf p}}
{(2\pi)^{2}}\frac{\exp(i{\bf p}\cdot{\bf r})}{{\bf p}^2+m^2}\nonumber\\
&
&+\frac{1}{m}({\hat z}\times{\bf\nabla})\left[\left({\bf V}\times{\hat z}\right)\cdot{\bf\nabla}\right]
\int\frac{d^{2}{\bf p}}{(2\pi)^{2}}\exp(i{\bf p}\cdot{\bf r})\left(\frac{1}{{\bf p}^2}-
\frac{1}{{\bf p}^2+m^2}\right)\ .
\end{eqnarray}

The first integral between brackets in the second line of the above expression is divergent.
In order to solve this problem we proceed as in references \cite{BaroneHidalgo1,DiracString1,DiracString4}, introducing a regulator parameter with dimension of mass, as follows
\begin{eqnarray}
\label{INTD1}
\int\frac{d^{2}{\bf p}}{(2\pi)^{2}}\frac{\exp(i{\bf p}\cdot{\bf r})}{{\bf p}^2}&=&\lim_{\mu\rightarrow0}
\int\frac{d^{2}{\bf p}}{(2\pi)^{2}}\frac{\exp(i{\bf p}\cdot{\bf r})}{{\bf p}^2+\mu^{2}}\nonumber\\
&=&\frac{1}{2\pi}\lim_{\mu\rightarrow0}\left[K_{0}\left(\mu\mid{\bf{r}}\mid\right)\right] \ .
\end{eqnarray}
Now, we use the fact that \cite{Arfken} $K_{0}(\mu\mid{\bf{r}}\mid)\stackrel{\mu\rightarrow0}{\rightarrow}
-\ln\left(\mu\mid{\bf{r}}\mid/2\right)-\gamma$, where $\gamma$ is the Euler constant, in order do handle the 
expression (\ref{INTD1}), thus
\begin{eqnarray}
\label{INTD2}
\int\frac{d^{2}{\bf p}}{(2\pi)^{2}}\frac{\exp(i{\bf p}\cdot{\bf r})}{{\bf p}^2}&=&
-\frac{1}{2\pi}\lim_{\mu\rightarrow0}\left[\ln\left(\frac{\mu\mid{\bf{r}}\mid}{2}\right)+\gamma\right]\nonumber\\
&=&-\frac{1}{2\pi}\lim_{\mu\rightarrow0}\left[\ln\left(\frac{\mu\mid{\bf{r}}\mid}{2}\right)+\gamma+\ln(\mu a_{0})-
\ln(\mu a_{0})\right]\nonumber\\
&=&-\frac{1}{2\pi}\left[\ln\left(\frac{\mid{\bf{r}}\mid}{a_{0}}\right)+\gamma-\ln 2+\lim_{\mu\rightarrow0}
\ln(\mu a_{0})\right]\nonumber\\
&=&-\frac{1}{2\pi}\ln\left(\frac{\mid{\bf{r}}\mid}{a_{0}}\right) \ .
\end{eqnarray}
Here, in the second line, we added and subtracted the quantity
$\ln\left(\mu a_{0}\right)$, where $a_{0}$ is an arbitrary constant 
with dimension of length. In the last line we neglected the terms 
that do not depend on the distance $\mid{\bf{r}}\mid$, since they do not 
contribute to the calculation of the 
electromagnetic field.

Substituting the result (\ref{INTD2}) into the Eq. (\ref{AKE}) and carrying out 
some manipulations, we arrive at
\begin{eqnarray}
\label{AKEEE}
{\bf A}\left({\bf{r}}\right)&=&\frac{m^{2}}{2\pi}
\Bigg[V^{0}\frac{K_{1}\left(mr\right)}{mr}({\bf r}\times{\hat z})\cr\cr
& &
+\left(\frac{K_{2}\left(mr\right)}{mr}-\frac{2}{(mr)^{3}}\right)
\frac{({\bf V}\times{\hat z})\cdot{\bf r}}{r}({\bf r}\times{\hat z})\cr\cr
&
&+\frac{1}{m}\left(\frac{K_{2}\left(mr\right)}{mr}-\frac{1}{(mr)^{2}}\right){\bf V}\Biggr] 
\ .
\end{eqnarray}

It is well known that the field strength $F^{\mu\nu}$ in Maxwell-Chern-Simons electrodynamics is given by
\begin{equation}
\label{matri1em}
F^{\mu\nu}=\bordermatrix{&        \cr
							& 0 \ \ &-E^{1} \ \ &-E^{2} \   \cr
              & E^{1} \ \ &0 \ \ &-B \       \cr
              & E^{2} \ \ &B \ \ &0 \   \cr}\   \ ,
\end{equation}
where the electric field ${\bf{E}}=\left(E^{1},E^{2}\right)$ has two components and the magnetic field $B$ has just one component.

Using the fact that $F^{\mu\nu}=\partial^{\mu}A^{\nu}-\partial^{\nu}A^{\mu}$, from the Eq's (\ref{matri1em}) and (\ref{AOE}), we obtain
\begin{eqnarray}
\label{VEF}
{\bf{E}}&=&-F^{01}{\hat{x}}-F^{02}{\hat{y}}\nonumber\\
&=&\frac{m^{2}}{2\pi}\Biggl[\left(\frac{K_{2}\left(mr\right)}{mr}\left({\bf V}\times{\hat z}\right)\cdot{\hat r}-V^{0}K_{1}\left(mr\right)\right){\hat{r}}\nonumber\\
&
&-\frac{K_{1}\left(mr\right)}{mr}\left({\bf V}\times{\hat z}\right)\Biggr]\ ,
\end{eqnarray}
where ${\hat{r}}$ is an unit vector pointing in the direction of the vector ${\bf{r}}$.

In the same way, from the Eq. (\ref{AKEEE}), the magnetic field produced by the topological
source reads
\begin{eqnarray}
\label{EMF}
B&=&-F^{12}\nonumber\\
&=&\frac{m^{2}}{2\pi}\left[K_{1}\left(mr\right)
\left({\bf V}\times{\hat z}\right)\cdot{\hat r}+V^{0}K_{0}\left(mr\right)\right] \ .
\end{eqnarray}

\begin{figure}[!h]
\centering
\includegraphics[scale=0.2]{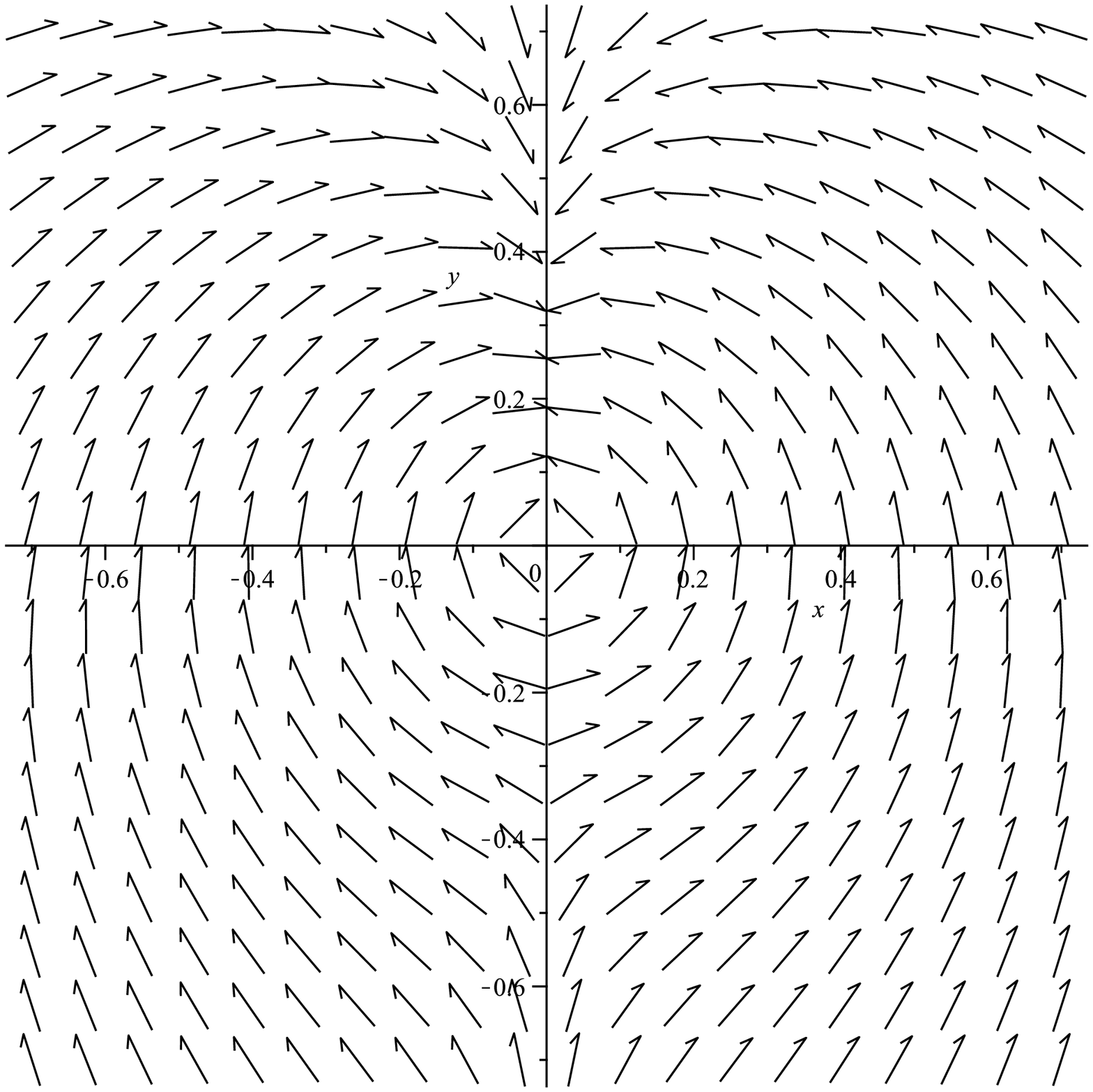}
\caption{Plot for the normalized electric field lines given by (\ref{VEF}).}
\label{linhasdecampo}
\end{figure}

The electric field (\ref{VEF}) exhibits a strong asymmetry. In can be seen in the graphic (\ref{linhasdecampo}), where we restrict to the specific case $V^{\mu}=V^{0}(1,1,0)$ and make a vector plot for the normalized electric field lines of (\ref{VEF}).

\begin{figure}[!h]
\centering
\includegraphics[scale=0.4]{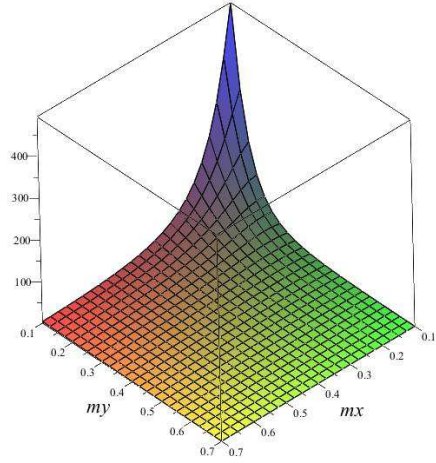}
\caption{Difference between the modulus of the electric field (\ref{VEF}) and the modulus of the magnetic field (\ref{EMF}) multiplied by $2\pi/m^2$.}
\label{diferencaentrecampos}
\end{figure}

In the Fig. (\ref{diferencaentrecampos}) we also take the same specific case where $V^{\mu}=V^{0}(1,1,0)$ and plot a graphic of the difference between the modulus of the electric field (\ref{VEF}) and the modulus of the magnetic one (\ref{EMF}). We can see that the graphic is always positive, what means that the electric field dominates over the magnetic one.  

By taking $V^{\mu}=\left(-\Phi,0\right)$ in the expressions (\ref{AKEEE}), (\ref{AOE}), (\ref{VEF}) and (\ref{EMF}), we obtain the electromagnetic
field produced by the Dirac point placed at origin as follows
\begin{eqnarray}
\label{EMFD}
A^{0}\left({\bf{r}}\right)=\frac{m\Phi}{2\pi}K_{0}\left(m \mid{\bf{r}}\mid\right)\ ,\ 
{\bf A}\left({\bf{r}}\right)=\frac{m\Phi}{2\pi}K_{1}\left(m \mid{\bf{r}}\mid\right){\hat\phi}\cr\cr
{\bf{E}}=\frac{m^{2}\Phi}{2\pi}K_{1}\left(m \mid{\bf{r}}\mid\right){\hat{r}} \ , \ 
B=-\frac{m^{2}\Phi}{2\pi}K_{0}\left(m \mid{\bf{r}}\mid\right) \ .
\end{eqnarray}

Notice that the second Eq. (\ref{EMFD}) is a vortex solution for the spatial part of the gauge field. 

For completeness, let us consider a point-like charge concentrated at origin,
\begin{equation}
\label{PCO}
J_{\nu}^{C}\left(y\right)=\sigma\eta_{\nu 0}\delta^{2}\left({\bf y}\right)\ .
\end{equation}
Following the same steps previously employed, it can be shown that
\begin{eqnarray}
\label{A0AKEMPC}
A^{0}\left({\bf{r}}\right)=\frac{\sigma}{2\pi}K_{0}\left(m \mid{\bf{r}}\mid\right) \ , \
{\bf A}\left({\bf{r}}\right)=\frac{\sigma}{2\pi}
K_{1}\left(m\mid{\bf{r}}\mid\right){\hat\phi} \ ,
\end{eqnarray}
What leads to
\begin{eqnarray}
\label{EBEM}
{\bf{E}}=\frac{m\sigma}{2\pi}K_{1}\left(m \mid{\bf{r}}\mid\right){\hat{r}} \ , \ 
B=-\frac{m\sigma}{2\pi}K_{0}\left(m \mid{\bf{r}}\mid\right) \ .
\end{eqnarray}

We highlight the similarity between the Eqs. (\ref{EMFD}), (\ref{A0AKEMPC}) and (\ref{EBEM}). If we identify $\sigma_{eff}=m\Phi$ in Eq. (\ref{EMFD}), we can see that the Dirac point behaves like a point-charge.

In the electromagnetic field configurations for the point-like particle (\ref{EBEM}) and for the Dirac point (\ref{EMFD}), we have a stronger contribution from the electric field in comparison with the magnetic one. The analysis for the topological source, with the fields (\ref{EMF}) and (\ref{VEF}), is more difficult, but for the specific case $V^{\mu}=V^{0}(1,1,0)$ the electric field also dominates in comparison with the magnetic one. This fact is expected for the case where we have two electric charges, but it is not expected for a setup with two Dirac points. In a magnetoelectric model, as the one given by (\ref{Lagrangian}), a charge density might induce field configurations where the electric field dominates over the magnetic ones. But for a current-type densities, like (\ref{topori}), it would be natural to expect that the magnetic field would dominate in comparison with the electric field, as occurs in topological insulators \cite{Science!323!1184,PRL!103!171601} and in some magnetoelectric (3+1)-dimensional media \cite{PRX!9!011011}, but not in the model (\ref{Lagrangian}). In our opinion, this is a rather remarkable result.

\begin{figure}[!h]
\centering
\includegraphics[scale=0.3]{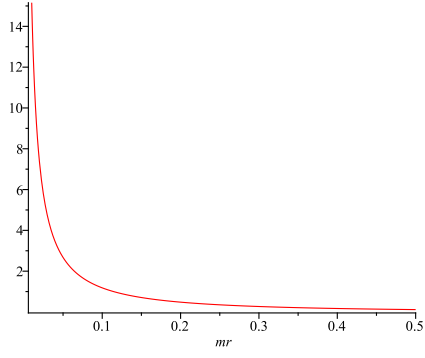}
\caption{Plot for the difference between the modulus of the electric and the magnetic fields divided by $m\sigma$, for the point-like charge, and $m^2\Phi$, for the Dirac point, of Eq's (\ref{EBEM}) and (\ref{EMFD}), respectively.}
\label{Eq45a}
\end{figure}

In figure (\ref{Eq45a}) we have a plot for the difference between the modulus of the electric and the magnetic fields divided by $m\sigma$, for the point-like charge, and $m^2\Phi$, for the Dirac point.   

\section{The Maxwell-Chern-Simons field in the presence of a conducting plate}
\label{MCSplaca}

In this section we consider the Maxwell-Chern-Simons electrodynamics with the presence
of a perfectly conducting plate. First of all, we must first establish what would be a perfectly conducting plate in this
theory. In Maxwell electrodynamics (in $3+1$ dimensions) the components of the Lorentz Force parallel to a
conducting surface must vanish on this surface. In  Maxwell-Chern-Simons electrodynamics ($2+1$ dimensions) the
coupling between the electromagnetic field and the charged particles exhibits the same structure as the corresponding
one in Maxwell theory (up to some peculiarities due to the dimensional reduction). So that in  Maxwell-Chern-Simons
electrodynamics we can describe the presence of a conducting surface by taking the components of the Lorentz force
parallel to the surface as being equal to zero. For a perfectly conducting plate lying on the plane $x^{2}=d$, it is
attained with the condition
\begin{equation}
\label{condition1}
n^{\mu  \ *}F_{\mu}|_{x^{2}=d}=0 \Rightarrow n^{\mu}\epsilon_{\mu}^{\ \nu\lambda}\partial_{\nu}A_{\lambda}|_{x^{2}=d}=0 \ ,
\end{equation}
where $n^{\mu}=(0,0,1)$ is the Minkowski vector normal to the plate and $ ^{*}F^{\mu}=(1/2)\epsilon^{\mu\nu\lambda}F_{\nu\lambda}$ is the dual to the field strength tensor. The condition (\ref{condition1}) asserts that the normal component of the dual field strength to the plate vanishes on the plane $x^{2}=d$.

Following a path integral approach, similar to what was done in references  \cite{LeeWick,Bordag,LHCFABplate,FABplate}, we write the functional generator as follows
\begin{eqnarray}
\label{fgen1}
Z_{C}\left[J\right]=\int {\cal{D}}A_{C} \ e^{i\int d^{3}x \ \cal{L}} \ ,
\end{eqnarray}
where the subscript $C$ indicates that we are integrating in all the field configurations which satisfy the conditions (\ref{condition1}). In an equivalent way, we can integrate out in all field configurations and introduce a delta functional, which is non-vanishing only for field configurations that satisfy the condition (\ref{condition1}), as follows
\begin{eqnarray}
\label{fgen2}
Z_{C}\left[J\right]=\int {\cal{D}}A \ \delta\left[n^{\mu  \ *}F_{\mu}|_{x^{2}=d}\right] \ e^{i\int d^{3}x \ \cal{L}} \ .
\end{eqnarray}

Now we use the functional Fourier representation
\begin{eqnarray}
\label{fgen3}
\delta\left[n^{\mu  \ *}F_{\mu}|_{x^{2}=d}\right]=\int {\cal{D}}B\exp\left[i
\int d^{3}x\ \delta\left(x^{2}-d\right)B\left(x_{\parallel}\right) {^{*}F_{2}\left(x\right)}\right] \ ,
\end{eqnarray}
where we defined the parallel coordinates to the plate $x_{\parallel}^{\mu}=\left(x^{0},x^{1},0\right)$ and $B\left(x_{\parallel}\right)$  is an auxiliary scalar field which depends only on the parallel coordinates.

Therefore, substituting (\ref{fgen3}) in (\ref{fgen2}) and then using (\ref{condition1}), the functional generator reads
\begin{eqnarray}
\label{fgen4}
Z_{C}\left[J\right]=\int {\cal{D}}A{\cal{D}}B\ e^{i\int d^{3}x \ \cal{L}}
\exp\left[-i\int d^{3}x\ \delta\left(x^{2}-d\right)A_{\beta}\left(x\right)
\epsilon_{2}^{\ \alpha\beta}\partial_{\alpha}B\left(x_{\parallel}\right)\right] \ .
\end{eqnarray}
It can noticed that the first exponential in (\ref{fgen4}) depends only on the gauge field $A^{\mu}$, but the second one contains a coupling among $A^{\mu}$ and $B$. In order to decouple $A^{\mu}$ and $B$, we perform the following translation
\begin{eqnarray}
\label{trans1}
A^{\beta}\left(x\right)\rightarrow A^{\beta}\left(x\right)+\int d^{3}y\ 
D^{\beta}_{\ \alpha}\left(x,y\right)\delta\left(y^{2}-d\right)\epsilon_{2}
^{\ \gamma\alpha}\partial_{\gamma}B\left(y_{\parallel}\right) \ ,
\end{eqnarray}
which has unitary jacobian and enables us to write (\ref{fgen4}) as follows
\begin{eqnarray}
\label{fgen5}
Z_{C}\left[J\right]=Z\left[J\right]{\bar{Z}}\left[J\right] \ ,
\end{eqnarray}
where $Z\left[J\right]$ is the usual functional generator  for the gauge field
\begin{eqnarray}
\label{fgen6}
Z\left[J\right]=\int{\cal{D}}A\ e^{i\int d^{3}x \ \cal{L}}
=Z\left[0\right]\exp\left[-\frac{i}{2}\int d^{3}x \ d^{3}y \ J^{\mu}\left(x\right)D_{\mu\nu}
\left(x,y\right)J^{\nu}\left(y\right)\right] \ ,
\end{eqnarray}
and ${\bar{Z}}\left[J\right]$ is a contribution due to the scalar field $B$  
\begin{eqnarray}
\label{fgen7}
{\bar{Z}}\left[J\right]=\int{\cal{D}}B\exp\left[i\int d^{3}x \ \delta
\left(x^{2}-d\right)I\left(x\right)B\left(x_{\parallel}\right)\right] \nonumber\\
\times\exp\left[-\frac{i}{2}\int d^{3}x \ d^{3}y \ \delta\left(x^{2}-d\right)
\delta\left(y^{2}-d\right)B\left(x_{\parallel}\right)W\left(x,y\right)
B\left(y_{\parallel}\right)\right] \ ,
\end{eqnarray}
where we defined
\begin{eqnarray}
\label{defi1}
I\left(x\right)=-\int d^{3}y \ \epsilon_{2}^{\ \gamma\alpha}
\left(\frac{\partial}{\partial x^{\gamma}}D_{\alpha\mu}\left(x,y\right)
\right)J^{\mu}\left(y\right) \ , \ W\left(x,y\right)=\epsilon_{2}^{\ \gamma\alpha}
\epsilon_{2}^{\ \beta\lambda}\frac{\partial^{2}D_{\lambda\alpha}\left(x,y\right)}
{\partial x^{\beta}\partial y^{\gamma}} \ .
\end{eqnarray}

Substituting (\ref{defi1}) and (\ref{propagator}) into (\ref{fgen7}), defining  the momentum parallel to the plate $p_{\parallel}^{\mu}=\left(p^{0},p^{1},0\right)$, the quantity $\Gamma=\sqrt{p_{\parallel}^{2}-m^{2}}$, and the parallel metric
\begin{eqnarray}
\label{etap}
\eta_{\parallel}^{\mu\nu}=\eta^{\mu\nu}-\eta_{\ 2}^{\mu}\eta^{\nu 2} \ ,
\end{eqnarray}
and using the fact that \cite{LeeWick} 
\begin{eqnarray}
\label{int}
\int \frac{dp^{2}}{2\pi}\frac{e^{i p^{2}\left(x^{2}-y^{2}\right)}}
{p^{\mu}p_{\mu}-m^{2}}=-\frac{i}{2\Gamma} \ e^{i\Gamma\mid x^{2}-y^{2}\mid} \ ,
\end{eqnarray}
where the integral above was computed in $dp^{2}$, with $p^{2}$ standing for the momentum 
component perpendicular to the plate, after an extensive calculation, we find
\begin{eqnarray}
\label{fgen8}
{\bar{Z}}\left[J\right]={\bar{Z}}\left[0\right]\exp\left[-\frac{i}{2}\int d^{3}x 
\ d^{3}y \ J^{\mu}\left(x\right){\bar{D}}_{\mu\nu}\left(x,y\right)J^{\nu}\left(y\right)\right] \ ,
\end{eqnarray}
where we defined the function
\begin{eqnarray}
\label{prop2}
{\bar{D}}_{\mu\nu}\left(x,y\right)&=&\frac{i}{2}\int \frac{d^{2}p_{\parallel}}
{\left(2\pi\right)^{2}} \ \frac{e^{-i p_{\parallel}\cdot\left(x_{\parallel}
-y_{\parallel}\right)}}{p_{\parallel}^{2}}\left(\epsilon_{2\gamma\mu}
p_{\parallel}^{\gamma}-im\eta_{2\mu}\right)\left(\epsilon_{2\beta\nu}
p_{\parallel}^{\beta}+im\eta_{2\nu}\right)
\nonumber\\
&
&\times\frac{e^{i\Gamma\left(\mid x^{2}-d\mid+\mid y^{2}-d\mid\right)}}{\Gamma} \ .
\end{eqnarray}

Substituting (\ref{fgen8}) and (\ref{fgen6}) in (\ref{fgen5}), the functional generator
of the Maxwell-Chern-Simons theory in the presence of a conducting plate reads
\begin{eqnarray}
\label{fgen9}
Z_{C}\left[J\right]=Z_{C}\left[0\right]\exp\left[-\frac{i}{2}
\int d^{3}x \ d^{3}y \ J^{\mu}\left(x\right)\left(D_{\mu\nu}
\left(x,y\right)+{\bar{D}}_{\mu\nu}\left(x,y\right)\right)J^{\nu}
\left(y\right)\right] \ .
\end{eqnarray}

From the Eq. (\ref{fgen9}), we can identify the gauge field propagator of the theory due to the presence of
a conducting plate as follows
\begin{eqnarray}
\label{prop3}
D_{C}^{\mu\nu}=D^{\mu\nu}\left(x,y\right)+{\bar{D}}^{\mu\nu}\left(x,y\right) \ .
\end{eqnarray}

The propagator (\ref{prop3}) is composed by the sum of the free propagator (\ref{propagator}) with the correction (\ref{prop2}) which accounts for the presence of the conducting plate. In the limit $m\rightarrow 0$ the propagator (\ref{prop2}) reduces to the same one as that found with the Maxwell electrodynamics in the presence of a conducting plate.

To check the validity of the results, we point out that the propagator (\ref{prop3}) under the boundary conditions is really a Green function for the problem, in the sense that it  satisfies the differential operator in Eq. (\ref{propO}), as follows
\begin{eqnarray}
\label{operplate}
\left[\eta_{\mu\nu}\partial_{\lambda}\partial^{\lambda}-m\epsilon_{\mu\nu\lambda}\partial^{\lambda}\right]D_{C}^{\nu\sigma}\left(x,y\right)=\eta_{\mu}^{\ \sigma}\delta^{3}\left(x-y\right) \ .
\end{eqnarray}

We can also consider the classical solutions for the gauge field obtained from (\ref{prop3})
\begin{eqnarray}
\label{pfield}
A^{\beta}(x)=\int d^{3}y \ D_{C}^{\beta\rho}(x,y) J_{\rho}(y) \ .
\end{eqnarray}
Substituting (\ref{pfield}) in Eq. (\ref{condition1}), the conducting plate condition reads
\begin{eqnarray}
\label{pfield11}
\int d^{3}y\left[\epsilon_{2\alpha\beta}\frac{\partial D_{C}^{\beta\rho}(x,y)}
{\partial x_{\alpha}}\right]J_{\rho}(y)|_{x^{2}=d}=0\cr\cr
\Rightarrow \epsilon_{2\alpha\beta}\frac{\partial D_{C}^{\beta\rho}(x,y)}
{\partial x_{\alpha}}\Big|_{x^{2}=d}=0\ ,
\end{eqnarray}
where the validity of the last line can be shown with the aid of (\ref{propagator}), 
(\ref{int}), (\ref{prop2}) and (\ref{prop3}).

\subsection{\label{II} Point-like charge and plate}

In this subsection we consider the interaction between a point-like charge and a conducting plate. It can be shown that the interaction energy between a conducting surface and an external source $J^{\nu}\left(x\right)$
is given by \cite{LeeWick,LHCFABplate,FAFEB}
\begin{eqnarray}
\label{energy}
{{E}}=\frac{1}{2T}\int d^{3}x \ d^{3}y \ J^{\mu}\left(x\right)
{\bar{D}}_{\mu\nu}\left(x,y\right)J^{\nu}\left(y\right) \ .
\end{eqnarray}

With no loss of generality and for simplicity, we choose
a point-like charge placed at the position ${\bf b}=\left(0,b\right)$. The external source is given by
\begin{eqnarray}
\label{source1}
J^{PC}_{\mu}\left(x\right)=q\eta^{0}_{\ \mu}\delta^{2}\left({\bf x}-{\bf b}\right) \ .
\end{eqnarray}

Substituting (\ref{source1}) and (\ref{prop2}) in (\ref{energy}) and carrying out the 
integrals in $d^{2}{\bf x}$, $d^{2}{\bf y}$, $dx^{0}$, $dp^{0}$, $dy^{0}$ and 
then performing some manipulations, we obtain
\begin{eqnarray}
\label{energy2}
{{E}}^{PC}=-\frac{q^{2}}{8\pi}\int d{\bf{p}}_{\parallel}\frac{e^{-2\mid b-d\mid\sqrt{{\bf{p}}
_{\parallel}^{2}+m^{2}}}}{\sqrt{{\bf{p}}_{\parallel}^{2}+m^{2}}} \ ,
\end{eqnarray}
where the superscript $PC$ means that we have the interaction energy between the conducting plate and the charge.

In order to solve the integral (\ref{energy2}) we use the fact that the integrand is an even function, we carry out the change 
in the integration variable $u=\sqrt{\frac{{\bf{p}}_{\parallel}^{2}}{m^{2}}+1}$ and use the fact that \cite{Gradshteyn}
\begin{eqnarray}
\label{int50}
\int_{1}^{\infty} du\frac{e^{-2m\mid b-d\mid u}}{\sqrt{u^{2}-1}}=K_{0}\left(2m\mid b-d\mid\right) \ ,
\end{eqnarray}
what leads to
\begin{eqnarray}
\label{energy3}
{{E}}^{PC}&=&-\frac{q^{2}}{4\pi}K_{0}\left(2m\mid b-d\mid\right) \ .
\end{eqnarray}

Now we fix the plate, use the fact that ${\bf b}=b{\hat y}$ and compute the force exerted on the point-like charge from expression (\ref{energy3}), as follows
\begin{eqnarray}
\label{force1}
{\bf F}^{PC}&=&-\frac{\partial{E}^{PC}}{\partial{\bf b}}=-\frac{\partial{E}^{PC}}{\partial b}{\hat y}=-\frac{m q^{2}}{2\pi}sgn(b-d)K_{1}\left(2m\mid b-d\mid\right){\hat y} \ ,
\end{eqnarray}
with $sgn$ standing for the sign function, defined by $sgn(x)=1 (x > 0), \ sgn(x)= -1 (x < 0), \  sgn(0)=0$, and where we used the fact that $b\neq d$, what means that the charge is not overlapped with the plate. The minus sign in (\ref{force1}) means that we have an attractive force. When the charge is on the right of the plate, $b>d$ and the force points in the $-{\hat y}$ direction. When the charge is on the left of the plate, $b<d$ and the force points in the ${\hat y}$ direction.

The interaction force (\ref{For1EM}) for the case where $\sigma_{1}=q$, $\sigma_{2}=-q$ and  
${\bf a}=2(b-d){\hat y}$ becomes equivalent to the one obtained in (\ref{force1}). Therefore, for this case the image 
method is valid for Maxwell-Chern-Simons electrodynamics for the conducting
plate condition (\ref{condition1}). A similar situation occurs in a Lorentz violation theory considered in \cite{LHCFABplate} where the image method is valid.

\subsection{\label{III} Topological source and plate}

Now we study the interaction energy between a topological field source and a conducting plate.

We consider the topological source placed at position  ${\bf b}=\left(0,b\right)$, as follows
\begin{eqnarray}
\label{source2}
J^{TP}_{\mu}\left(x\right)=\epsilon_{\mu}^{\ \alpha\beta}V_{\alpha}
\partial_{\beta}\delta^{2}\left({\bf x}-{\bf b}\right) \ .
\end{eqnarray}

Substituting (\ref{source2}) and (\ref{prop2}) in (\ref{energy}), following the same steps employed in the previous sections, performing the same changes in the integration variable employed in subsection \ref{II} and using (\ref{int50}) and the fact that \cite{Gradshteyn}
\begin{eqnarray}
\label{int60}
\int_{1}^{\infty} du \sqrt{u^{2}-1} \ e^{-2m\mid b-d\mid u}=\frac{K_{1}\left(2m\mid b-d\mid\right)}{2m\mid b-d\mid}\ ,
\int_{1}^{\infty} du \  u \ \frac{e^{-2m\mid b-d\mid u}}
{\sqrt{u^{2}-1}}=K_{1}\left(2m\mid b-d\mid\right) \ ,
\end{eqnarray}
we arrive at
\begin{eqnarray}
\label{energyTP}
E^{TP}&=&-\frac{m^{2}}{4\pi}\Biggl[\left(\frac{{\bf {V}}^{2}}{2m\mid b-d\mid}+2V^{0}V^{1}\frac{b-d}{\mid b-d\mid}\right)
K_{1}\left(2m\mid b-d\mid\right)\nonumber\\
&
&+\left(\left(V^{0}\right)^{2}+\left(V^{1}\right)^{2}\right)
K_{0}\left(2m\mid b-d\mid\right)\Biggr] \ .
\end{eqnarray}
where the superscript $TP$ means that we have the interaction between the topological source and the plate.

By fixing the plate (fixing the variable $a$), taking into account that ${\bf b}=b{\hat y}$, and using the fact that 
\begin{equation}
\frac{K_{0}(2x)}{x}+\frac{K_{1}(2x)}{x^2}=\frac{K_{2}(2x)}{x}\ ,
\end{equation}
we have the force acting on the topological source, as follows
\begin{eqnarray}
\label{force2}
{\bf F}^{TP}&=&-\frac{\partial E^{TP}}{\partial{\bf b}}=-\frac{\partial E^{TP}}{\partial b}{\hat y}=\nonumber\\
&=&-\frac{m^{2}}{4\pi}\Bigg[sgn\left(b-d\right){\bf V}^{2}\frac{K_{2}\left(2m\mid b-d\mid\right)}{\mid b-d\mid} \nonumber\\
&\ &-2V^{0}V^{1}m\left[K_{0}\left(2m\mid b-d\mid\right)+K_{1}\left(2m\mid b-d\mid\right)\right]\nonumber\\
&\ &+sgn\left(b-d\right)\left(\left(V^{0}\right)^{2}+\left(V^{1}\right)^{2}\right)2mK_{1}\left(2m\mid b-d\mid\right)\Bigg]{\hat y}
\end{eqnarray}
which is always perpendicular to the plate.

Expression (\ref{force2}) exhibits some interesting features. The first one is the fact that force (\ref{force2}) does not satisfy the image method. This point must be analysed carefully.

In a mirror reflection, the normal component (with respect to the mirror) of a polar vector  inverts its sign, while its parallel components to the mirror remain unchanged. For an axial vector, we have the opposite situation; in a mirror reflection, the normal component to the mirror remain unchanged and their parallel components are inverted. In our case, ${\bf V}$ is a 2-dimensional pseudo-vector (axial vector) and we have a mirror on the plane $x^{2}=d$. So that, the normal component to the mirror of ${\bf V}$ is $V^{2}$ and we have just one parallel component to the mirror, $V^{1}$.

In what concerns the temporal component of the axial vector; for a boost with speed ${\bf v}$, which is a polar vector, the product ${\bf v}\cdot{\bf V}$ exhibits an overall sign inversion under a reflection on a mirror, because ${\bf V}$ is a pseudo-vector. Taking into account that a boost for the $0$ component of $V^{\mu}$ is given by $(V^{0}-{\bf v}\cdot{\bf V})$, we conclude that $V^{0}$ must have its sign inverte on a reflection, in order to assure a consistent boosts for the pseudo Minkowski vector.  

To evaluate Eq. (\ref{fortptp}) for ${\bf{a}}=2(b-d){\hat y}$ with a supposed image of the topological source described by $U^{\mu}=(U^{0},{\bf U})$, we must consider the image condition for a pseudo Minkowski vector $U^{1}=V^{1},U^{2}=-V^{2},U^{0}=-V^{0}$. So that, one can show that expression (\ref{fortptp}), with the image topological source, is not equal to the force (\ref{force2}). We could also consider a Minkowski vector, by taking $U^{1}=-V^{1},U^{2}=V^{2},U^{0}=V^{0}$, the conclusion would be the same.

Thereby, the topological source does not satisfy the image method. As far as the authors know, the image method for stationary sources was thought to be invalid only in quadratic theories  with higher order derivatives \cite{LeeWick}.

Taking $V^{\mu}=-\Phi\eta^{\mu0}$ in the expression (\ref{force2}), we obtain the interaction force that the conducting plate exerts on the Dirac point, as follows
\begin{eqnarray}
\label{dpforce}
{\bf{{{F}}}}^{TP}\left({\bf{V}}=0,V^{0}=-\Phi\right)={\bf{{{F}}}}^{DP}=-\frac{m^{3}\Phi^{2}}{2\pi}K_{1}\left(2m\mid b-d\mid\right)sgn(b-a){\hat y} \ , 
\end{eqnarray}
where the superscript $DP$ means the interaction between a Dirac point and the plate. In the limit $m\rightarrow 0$,
this force vanishes. If we identify  $q_{eff}=m\Phi$ in Eq. (\ref{dpforce}), from the Eq. (\ref{force1}) we can see that the Dirac point behaves like a point-charge when it interacts with the conducting plate.

For the case where ${\bf{a}}=2(b-d){\hat y}$, $\Phi_{1}=-\Phi_{2}=\Phi$ (since $U^{0}=-V^{0}$), the Eq. (\ref{For2EM}) turns out to be equivalent to the Eq. (\ref{dpforce}). So, it is interesting to notice that the image method is valid for the Dirac-point for the conducting plate condition (\ref{condition1}). This result was already expected, since the Dirac-point behaves as point-like charges as discussed in the Section (\ref{fontes}).

For the standard Maxwell electrodynamics the interaction force that the plate exerts on the topological source can be obtained by taking the limit $m\rightarrow 0$ in Eq. (\ref{force2}). The result reads
\begin{eqnarray}
\label{tpforceM}
{\bf{{{F}}}}^{TP}\left(m=0\right)={\bf{{{F}}}}^{TP}_{M}=-\frac{{\bf{V}}^{2}}{8\pi\mid{b-d}\mid^{3}}sgn(b-d){\hat y} \ .
\end{eqnarray}

Proceeding in the same way as in the previous cases, it is simple to verify from Eq. (\ref{fortptpm}) that the image method is valid in Maxwell electrodynamics for the topological field source. This result was expected because, as discussed before, in the masless case just the spatial components $\bf V$ are relevant, and it corresponds to a typical electric dipole in $2+1$ dimensions.

Other interesting feature of the force (\ref{force2}) is the fact that it does not exhibit symmetry under spatial reflection on the mirror. Let us consider two setups composed by the plate and a topological source. In the first setup, a source $1$, given by $V^{\mu}$ is placed at position $b_{1}=d+s$ ($s>0$). In the second setup, we take a source $2$, with the same (pseudo) vector $V^{\mu}$ placed just at the reflected position $b_{2}=d-s$. For each case, the corresponding force (\ref{force2}) is, respectively.
\begin{eqnarray}
\label{TP1}
{\bf{{{F}}}}^{TP}_{1}
&=&-\frac{m^{2}}{4\pi}\Bigg[{\bf V}^{2}\frac{K_{2}\left(2ms\right)}{s} \nonumber\\
&\ &+\left(\left(V^{0}\right)^{2}+\left(V^{1}\right)^{2}\right)2mK_{1}\left(2ms\right)\nonumber\\
&\ &-2V^{0}V^{1}m\left[K_{0}\left(2ms\right)+K_{1}\left(2ms\right)\right]\Bigg]{\hat y}\ , \\
\label{TP2}
{\bf{{{F}}}}^{TP}_{2}
&=&-\frac{m^{2}}{4\pi}\Bigg[-{\bf V}^{2}\frac{K_{2}\left(2ms\right)}{s} \nonumber\\
&\ &-\left(\left(V^{0}\right)^{2}+\left(V^{1}\right)^{2}\right)2mK_{1}\left(2ms\right)\nonumber\\
&\ &-2V^{0}V^{1}m\left[K_{0}\left(2ms\right)+K_{1}\left(2ms\right)\right]\Bigg]{\hat y}\ .
\end{eqnarray}

Notice that the first terms on the right hand sides of Eq's (\ref{TP1}) and (\ref{TP2}) exhibit spatial reflection symmetry   on the mirror, but the third terms on these equations does not exhibit it. It is a feature of the topological source due to the break of spatial reflection on the mirror. This effect has no counterpart in the usual Maxwell electrodynamics, what can be directly verified by taking the limit $m\rightarrow 0$ in Eq. (\ref{force2}) or, alternatively, in Eq's (\ref{TP1}) and (\ref{TP2}), where we recover expressions with spacial reflection symmetry   on the mirror. This effect is also absent when we consider a Dirac point, where ${\bf V}=0$. Therefore, this asymmetry brings out only if $V^{0}\not=0$ and ${\bf V}\not={\bf 0}$.

When we fix the distance between the topological source and the plate, from Eq. (\ref{energyTP}), we see that the whole system undergoes a torque given by its  orientation with respect to the vector ${\bf{V}}$. In order to calculate this torque, we define as $0 \leq \alpha \leq \pi$ the angle between the normal to the plate and the vector ${\bf{V}}$, in such a way that
\begin{equation}
\label{vpvp}
\left(V^{2}\right)^{2}={\bf{V}}^{2}\cos^{2}\left(\alpha\right), \ \ \ \left(V^{1}\right)^{2}={\bf{V}}^{2}\sin^{2}\left(\alpha\right) \ ,
\end{equation}
and we can rewrite Eq. (\ref{energyTP}) as function of the angle $\alpha$, as follows
\begin{eqnarray}
\label{eneralpha}
E^{TP}\left(\alpha\right)=-\frac{m^{2}}{4\pi}\Biggl[\left(\frac{{\bf{V}}^{2}}{2m\mid b-d\mid}
+2V^{0}\sqrt{{\bf V}^{2}}\sin(\alpha)\frac{b-d}{|b-d|}\right)K_{1}\left(2m\mid b-d\mid\right)\nonumber\\
+\left((V^{0})^{2}+{\bf{V}}^{2}\sin^{2}\left(\alpha\right)\right)K_{0}\left(2m\mid b-d\mid\right)\Biggr] \ .
\end{eqnarray}
Just for simplicity, taking $V^{0}=0$, from the Eq. (\ref{eneralpha}) the torque reads
\begin{equation}
\label{eneralpha2}
\tau^{TP}=-\frac{\partial }{\partial\alpha}E^{TP}\left(V^{0}=0,\alpha\right)=\frac{m^{2}{\bf{V}}^{2}}{4\pi}\sin\left(2\alpha\right)K_{0}\left(2m\mid b-d\mid\right) \ .
\end{equation}
If $\alpha=0,\pi/2,\pi$ the torque vanishes, when $\alpha=\pi/4$, it exhibits a maximum value. In the limit $m\rightarrow 0$, the torque is equal to zero.

\section{Conclusions}
\label{conclusoes}

In this paper some new aspects of the so called Maxwell-Chern-Simons electrodynamics due to the interactions between stationary point-like sources as well as the presence of a perfectly conducting plate have been investigated. Specifically, in addition to the point-like charges, we proposed two new kinds of  point-like sources, which we called topological source and Dirac point, and we considered effects which emerged of interactions between these sources. We showed that the Dirac point behaves similarly to point-like charges and the topological source induces the presence of torques in all the setups considered. 

We studied the field configurations produced by the point-like charge, the Dirac point and the topological source, and showed that the Dirac point leads to vortex configurations for the gauge field. 

The propagator for the gauge field due to the presence of a conducting plate and the interaction forces between the plate and the point-like sources were computed. The conclusion is that the image method is valid for the point-like charges as well as for Dirac points. On the other hand, for  topological sources we showed that the image method is not valid. 

We also have shown that the interaction force between the plate and the topological source does not exhibit spatial reflection symmetry   on the mirror.  This feature is due to the spatial asymmetry imposed by the presence of the topological source. Another interesting fact is the emergence of a torque acting on the plate when it interacts with a topological source.

\begin{acknowledgments}
For financial support, L. H. C. Borges thanks to S\~ao Paulo Research Foundation (FAPESP) under the grant 2016/11137-5, F. A. Barone thanks to CNPq (Brazilian agency) under the grants 311514/2015-4 and 313978/2018-2, C. C. H. Ribeiro thanks to CAPES and FAPEMIG  (Brazilian agencies) and H. L. Oliveira thanks to CAPES. The authors would like to thank J. A. Helay\"el-Neto for very valuable comments and suggestions. 
\end{acknowledgments}

\end{document}